# Magnetic properties of BiFeO$_3$ – BaTiO$_3$ ceramics in the morphotropic phase boundary: a role of crystal structure and structural parameters

D.V. Karpinsky[1,2,3], M.V. Silibin[2,4,5], D.V. Zhaludkevich[1*], S.I. Latushka[1], A.V.Sysa[2,4], V.V. Sikolenko[6, 6a, 6b], A.L. Zhaludkevich[1], V.A. Khomchenko[7], A. Franz[8], K. Mazeika[9], D. Baltrunas[9], A. Kareiva[10]

[1] *Scientific-Practical Materials Research Centre of NAS of Belarus, 220072 Minsk, Belarus*
[2] *National Research University of Electronic Technology "MIET", 124498 Zelenograd, Moscow, Russia*
[3] *Scientific and Educational Center "Nanotechnology", South Ural State University, 454080, Chelyabinsk, Russia*
[4] *Scientific-Manufacturing Complex "Technological Centre", 124498 Zelenograd, Moscow, Russia*
[5] *Institute for Bionic Technologies and Engineering, I.M. Sechenov First Moscow State Medical University, 119991 Moscow, Russia*
[6] *REC "Functional nanomaterials", Immanuel Kant Baltic University, 236041 Kaliningrad, Russia*
[6a] *Joint Institute for Nuclear Research, 141980 Dubna, Russia*
[6b] *Karlsruhe Institute of Technology, 76131 Karlsruhe, Germany*
[7] *CFisUC, Department of Physics, University of Coimbra, 3004-516 Coimbra, Portugal*
[8] *Helmholtz-Zentrum Berlin für Materialien und Energie, 14109 Berlin, Germany*
[9] *State Research Institute Center for Physical Sciences and Technology, LT-02300 Vilnius, Lithuania*
[10] *Institute of Chemistry, Vilnius University, Naugarduko 24, LT-03225 Vilnius, Lithuania*

ABSTRACT

A correlation between the crystal structure and magnetic properties of system (1-x)BiFeO$_3$ – (x)BaTiO$_3$ with compounds across the morphotropic phase boundary was studied using X-ray and neutron diffraction, magnetometry, and Mössbauer spectroscopy measurements. Increase in the dopants content leads to the structural transition from the rhombohedral phase to the cubic phase via a formation of the two-phase region (0.2 < x < 0.33), wherein the magnetic structure changes from the modulated G-type antiferromagnetic to the collinear antiferromagnetic via a stabilization of the non-collinear antiferromagnetic phase with non-zero remanent magnetization. The value of magnetic moment calculated per iron ion based on the Mössbauer and neutron diffraction data decreases from m ≈ 4.4 $\mu_B$ for the compound with x=0.25 to m=3.2 $\mu_B$ for the compound with x=0.35 testifying a dominance of 3+ oxidation state of the iron ions. Increase in the amount of the cubic phase leads to a reduction in the remanent magnetization from 0.02 emu/g for the compounds with the dominant rhombohedral phase (x < 0.27) down to about 0.001 emu/g for the compounds with dominant cubic structure (x ≥ 0.27). Rapid decrease in the remanent magnetization observed in the compounds across the phase coexistence region points at no direct correlation between the type of structural distortion and non-zero remanent magnetization, while the oxygen octahedra tilting is the key factor determining the presence of non-zero remanent magnetization.

1. INTRODUCTION

During the last decades, the materials having multiferroic properties attract great attention of the scientific community due to the fundamental aspects and perspectives of practical applications [1-9]. Bismuth ferrite and related complex oxides are considered to be the most popular multiferroic materials exhibiting magnetoelectric coupling at room temperature [10-16]. The most intriguing properties of bismuth ferrite based oxides were observed for the compounds near the morphotropic phase boundaries. The compounds with compositionally-induced two- or three-phase structural state are of particular importance for understanding the background of the phase boundary region as well as for analyzing the factors determining the temperature- and concentration ranges peculiar to different structural phases. Simultaneous Bi-/Fe-site substitution in the parent $BiFeO_3$ with alkali-earth/transition metal ions allows to create compounds within the morphotropic phase boundary region and to control their electric dipole and magnetic long range orderings [17-24]. The long range ferroic orders can be controllably frustrated using chemical doping to stabilize the so called *multi glass* state which is characterized by intriguing functional properties [25, 26]. Frustration of the long range ferroic order accompanied by the reduced structural stability characteristic of the compounds near the morphotropic phase boundary provides a new insight on the perspectives of the $BiFeO_3$-based multiferroic materials.

The rare-earth doped $BiFeO_3$ compounds are characterized by a direct correlation between the type of lattice distortion and a remanent magnetization [27-29]. A formation of the orthorhombic phase in the phase boundary region of $Bi_{1-x}RE_xFeO_3$ ceramics leads to a release of remanent magnetization which is cancelled in the compounds with single phase rhombohedral structure because of the spatial modulation of the magnetic structure [1, 30]. Analysis of literature data available for the co-doped compounds $Bi_{1-x}A_xFe_{1-x}Ti(M)_xO_3$ (A - alkali-earth ions, M - transition metal element) testifies that a correlation between the crystal structure, viz. the type of structural distortion and the presence of non-zero remanent magnetization is still under discussion [31-34]. In spite of the available data, there is no reliable model which describes a correlation between the symmetry, structural parameters and magnetic properties of co-doped $BiFeO_3$ compounds across the phase boundary regions.

The objective of the present study is focused on a clarification of the factors which cause a formation of non-collinear antiferromagnetic structure and associated non-zero remanent magnetization in $BiFeO_3$-based compounds. The present study is focused on the Ba- and Ti- doped $BiFeO_3$ ceramics as these compounds have one of the widest (~15 %) concentration driven phase transition among the co-doped $BiFeO_3$ systems. Presented data clarify the correlation between the remanent magnetization, the type of structural distortion and the structural parameters as the chemical bond lengths, angles and oxygen octahedra tilting and rotation in the compounds $Bi_{1-x}Ba_xFe_{1-x}Ti_xO_3$ across the rhombohedral - cubic phase transition. A relation between the type of lattice distortion, structural parameters and magnetic properties is analyzed for the ceramics $(1-x)BiFeO_3$ – (x)BaTiO$_3$ with x < 0.40 depending on the dopant content and temperature. Careful analysis performed by the authors has allowed to provide a novel model describing magnetic properties, viz. the presence of non-zero remanent magnetization in BiFeO$_3$ compounds co-doped in A- and B- perovskite positions; the structural parameters associated with oxygen octahedra are determined to be the main factors causing a formation of non-zero remanent magnetization.

2. EXPERIMENTAL

Ceramic samples of the series (1-x)BiFeO$_3$ – (x)BaTiO$_3$ (0.15 < x < 0.40) were prepared by the two stage solid-state reaction technique [20]. High-purity (≥ 99.5%, Alfa Aesar) simple oxides Bi$_2$O$_3$, Fe$_2$O$_3$, TiO$_2$ and carbonite BaCO$_3$ taken in stoichiometric ratio were thoroughly mixed in ethanol (96%) medium using a ball mill Retsch PM-100. Stainless steel balls (⌀5 mm) were used as durable balls which do not provide any notable effect on the phase purity and magnetic properties of the compounds assuming the used mixing conditions (300 rpm, 30 min). Preliminary synthesis of the samples uniaxially pressed into tablets (diameter is 10 mm) was performed at 900°C for 3 h. After intermittent grinding, the samples were finally synthesized at temperatures 910 - 945°C for 12 h followed by a cooling with a rate of about 50°C/h (the synthesis temperature was gradually increased by ∼ 5°C per each 5% of the dopants concentration increase). After synthesis, the samples were cooled down to room temperature with a cooling rate of 100 °C/h. Phase purity and crystal structure of the compounds were attested based on X-ray diffraction (XRD) data obtained using a Bruker D8 Advance diffractometer with Cu-Kα radiation and recorded in the 2Thetta range 20 - 80° with a step of 0.02°. Crystal and magnetic structures of the compounds were analyzed using the neutron powder diffraction (NPD) data obtained at the high-resolution neutron powder diffractometer FIREPOD (λ=1.7977Å, E9 instrument, HZB) equipped with a dedicated cryofurnace [35]. XRD and NPD data were refined by the Rietveld method using the FullProf software [36]. The Mössbauer measurements were performed in transmission geometry using a spectrometer Wissenschaftliche Elektronik GmbH with a $^{57}$Co(Rh) source. The spectra were recorded in the temperature range 10 – 550 K using a closed cycle He cryostat (Advanced Research Systems) and special Mössbauer furnace. The spectra were fitted to subspectra (sextets), hyperfine field distributions and doublet/singlet using the WinNormos software, the isomer shifts were related to the α-Fe standard. Magnetization measurements were performed in the temperature range 5 – 310K in magnetic fields up to 14 T using Physical Properties Measurement System from Cryogenic Ltd.

3. RESULTS AND DISCUSSION

The phase purity of the compounds (1-x)BiFeO$_3$ – (x)BaTiO$_3$ is confirmed by the results of the X-ray and neutron diffraction measurements (Fig. 1, 2). The diffraction data testify a negligible (if any) presence of impurity phases and thus assure an intrinsic character of the modification of the magnetic properties and changes in the crystal structure of the compounds taking place with the dopant content and temperature change. The crystal structure of the compounds over the concentration range 0 ≤ x ≤ 0.4 undergoes the consecutive concentration-driven phase transition from the polar active rhombohedral structure (space group $R3c$) to the cubic structure (s.g. $Pm\bar{3}m$) via a formation of the two-phase concentration region. The phase coexistence range is ascribed to the concentration region $0.20 < x < 0.33$.

In the mentioned phase coexistence region, the value of rhombohedral distortion gradually decreases as evidenced by the parameters presented in the Table #1. In particular, an increase in the dopant content leads to a notable increase of the $a_R$-parameter, whereas the $c_R$-parameter remains nearly constant. Accordingly, the c/a ratio decreases thus specifying a reduction of rhombohedral distortion. The coordinates of the ions determined based on the results of X-ray and neutron diffraction measurements also testify regular shift toward more symmetric positions. In particular, the Fe and Ti ions located in the 6a Wyckoff position gradually move to the symmetric position (0; 0; 0.25), while the oxygen ions occupying the 18b Wyckoff position move toward the symmetric position (0.5; 0; 1). Gradual shift of the ionic coordinates occurred with an increase in the dopant concentration leads to an equalization of the chemical bond lengths Fe – O and increase of the chemical bond angles Fe – O – Fe to 180° in the compounds with x > 0.3 (Table 1).

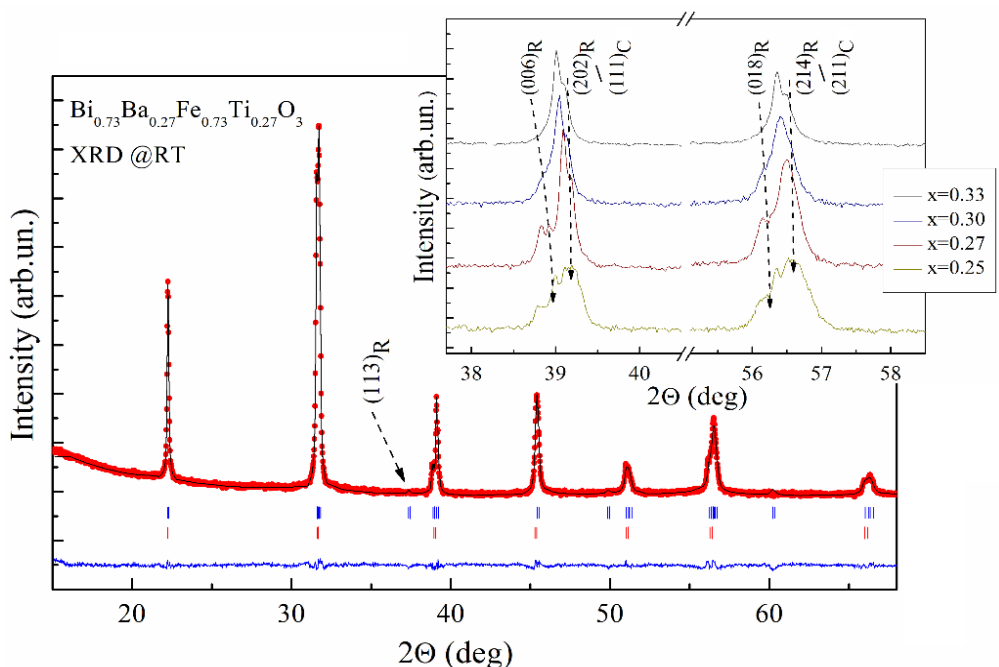

Figure 1. Room-temperature XRD pattern of $Bi_{0.73}Ba_{0.27}Fe_{0.73}Ti_{0.27}O_3$ refined in the two phase model (upper row of vertical ticks denote Bragg positions of the rhombohedral phase). The inset shows a concentration-driven evolution of the diffraction peaks ascribed to the rhombohedral and cubic phases.

The mentioned changes in the structural parameters are associated with a modification in the intensity and the positions of the reflections specific for the rhombohedral phase. The reflection $(113)_R$ associated with the $a^-a^-a^-$ antiphase tilting of oxygen octahedra gradually disappears with the dopant concentration (Fig. 1); the reflections $(202)_R$ and $(006)_R$ which characterize the ratio $c/a$ show a merging with an increase in the dopant content, thus pointing at a decrease of the rhombohedral distortions (Fig. 1). The diffraction patterns recorded for the compounds with x > 0.3 were successfully refined assuming a single phase cubic structure. It should be noted that the diffraction peaks observed for the compounds 0.20 ≤ x ≤ 0.30 are characterized by certain widening which is common for compounds having two phase structural state. The XRD data obtained for the compound with x = 0.33 (Fig. 1, inset) confirm a completion of the transition to the single phase cubic structure accompanied by a gradual increase in the unit cell volume (Table 1). It should be noted that the compounds 0.30 ≤ x ≤ 0.35 are also characterized by small widening of the reflections as compared to those obtained for the compound with x = 0.4 which can point at certain structural inhomogeneity of the compounds, most probably caused by a presence of nanoscale clusters of the rhombohedral phase which cannot be confidently detected by the diffraction methods.

Table 1. Unit cell parameters *, selected interatomic distances and angles, tilting angle, coordinates (Fe and O ions in cubic $Pm\bar{3}m$ phase are located in the positions Fe:1b(0.5;0.5;0.5) and O:3c(0;0.5;0.5); in rhombohedral phase $R3c$ – Fe:6a(0;0;z) and O:18b(x;y;z)), magnetic moments calculated for the compounds $Bi_{1-x}Ba_xFe_{1-x}Ti_xO_3$ with x=0.15 - 0.40 using neutron diffraction data obtained at T=300 K.

| x, % | phase ratio | $a_R$, Å | $c_R$, Å | Fe-O-Fe, ° | Fe-O, Å | ω, ° | Fe(Ti)$_R$: (0;0;z) | O$_R$: (x;y;z) | m, $\mu_B$ |
|---|---|---|---|---|---|---|---|---|---|
| 15 | $R3c$ | 3.9611(6) | 4.0103(3) | 160.4(3) | 2.084/1.952 | 14.9 | 0.2249 | 0.463/0.0166/0.9586 | 4.64 |
| 20 | $R3c$ (90%) $Pm\bar{3}m$ (10%) | 3.9727(5) | 4.0102(5) | 168.3(7) | 2.052/1.958 | 11.4 | 0.2211 | 0.469/0.0108/0.9739 | 4.57 |
|  |  | 3.9796(5) |  | 180.0 | 1.989 | - | - | - |  |
| 25 | $R3c$ (55%) $Pm\bar{3}m$ (45%) | 3.9736(5) | 4.0077(6) | 172.6(2) | 2.043/1.959 | 7.45 | 0.2291 | 0.480/0.0070/0.9803 | 4.37 |
|  |  | 3.9836(8) |  | 180.0 | 1.992 | - | - | - |  |
| 27 | $R3c$ (35%) $Pm\bar{3}m$ (65%) | 3.9811(5) | 4.0023(6) | 176.2(2) | 2.024/1.971 | 3.51 | 0.2393 | 0.491/0.0039/0.9921 | 3.94 |
|  |  | 3.9852(4) |  | 180.0 | 1.993 | - | - | - |  |
| 30 | $R3c$ (10%) $Pm\bar{3}m$ (90%) | 3.9833(4) | 4.0020(5) | 177.1(9) | 2.021/1.973 | 2.81 | 0.2409 | 0.493/0.0031/0.9941 | 3.55 |
|  |  | 3.9869(4) |  | 180.0 | 1.994 | - | - | - |  |
| 33 | $Pm\bar{3}m$ | 3.9911(7) | -/- |  | 1.995 | - | - | - | 3.34 |
| 35 | $Pm\bar{3}m$ | 3.9923(5) | -/- |  | 1.996 | - | - | - | 3.23 |
| 40 | $Pm\bar{3}m$ | 3.9925(4) | -/- |  | 1.996 | - | - | - | 3.12 |

* unit cell parameters presented in reduced form

The concentration-driven modification of the structural parameters estimated based on the results of the diffraction measurements governs the changes in the magnetic properties of the compounds. Dopant increase leads to an equalization of the chemical bond lengths Fe(Ti) - O as well as a modification of the corresponding angle Fe(Ti) – O - Fe(Ti) which gradually increases from 160.4° for compound with x=0.15 to about 176.2° for x=0.27 and further towards 180° for cubic compounds (Table 1). It should be noted that the bond angle Fe(Ti) – O - Fe(Ti) and the angle ω determine the oxygen octahedra rotation and tilt [37], which are the critical parameters affecting the remanent magnetization driven by the antisymmetric Dzyaloshinskii-Moriya interaction between neighboring iron ions [38-40].

Analysis of the neutron diffraction data allowed a more precise determination of the structural parameters calculated based on the XRD data as well as the refinement of magnetic structure of the compounds (Fig. 2). The magnetic moment calculated using the NPD data decreases from about 4.64 $\mu_B$ per iron ion at room temperature for the compound with x = 0.15 down to about 3.12 $\mu_B$ for compound with x=0.4, testifying a diamagnetic dilution of the magnetic sublattice formed by iron ions $Fe^{3+}$ by chemical doping with nonmagnetic $Ti^{4+}$ ions. The magnitudes of the magnetic moments calculated based on the NPD results are in accordance with the "spin only" value ascribed to the iron ions in 3+ oxidation state (electronic configuration $t_{2g}^5 e_g^0$) while a decrease in the magnetic moment is associated with the concentration-driven suppression of the long range antiferromagnetic order formed by iron ions.

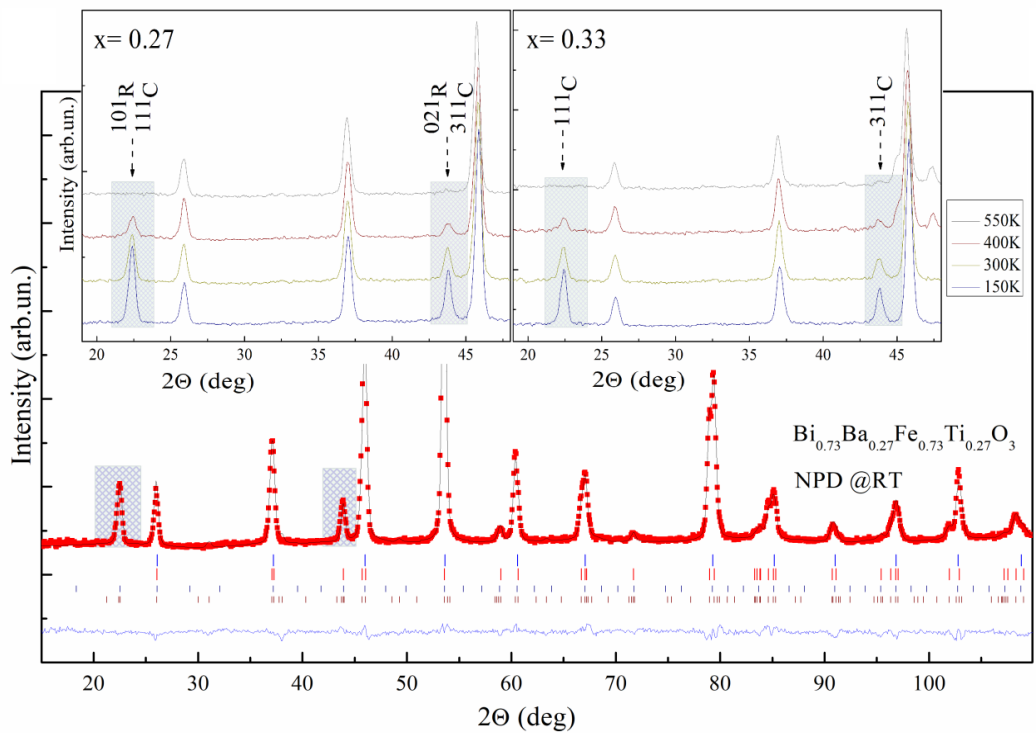

Figure 2. Room-temperature NPD pattern of $Bi_{0.73}Ba_{0.27}Fe_{0.73}Ti_{0.27}O_3$ refined in the two phase model. Two upper rows of vertical ticks denote Bragg positions of the cubic and rhombohedral phases respectively, two bottom rows – magnetic phases associated with related structural phases. The insets shows a temperature-driven evolution of the diffraction peaks ascribed to the crystal and magnetic structures (magentic reflections are highligted and indexed).

The room-temperature neutron diffraction data confirm the formation of the G-type antiferromagnetic structure in the compounds under study (Fig. 2). The temperature dependent data point at a gradual decrease of the Neel point from $T_N \approx 640$ K for initial undoped compound to $T_N \approx 420$ K for compound with x=0.4 (Fig. S1). It should be noted that analysis of the NPD data did not reveal any notable ferromagnetic component while magnetization measurements shows non-zero remanent magnetization in the compounds. Moreover, the remanent magnetization which is caused by the antisymmetric Dzyaloshinskii-Moriya exchange interaction allowed in the rhombohedral phase rapidly nullifies in the compounds having notable amount of the rhombohedral phase, thus implying a nontrivial correlation between the magnetization and the structural parameters of the compounds associated with the phase boundary region.

*Magnetization measurements*

The magnetization measurements have allowed to itemize the changes occurred in the magnetic structure depending on the dopant content and temperature. The M(H) dependences obtained at T ~ 5 K and room temperature show nearly linear character of the magnetization curves (Fig. 3), thus denoting the antiferromagnetic structure of the compounds under study, while temperature dependent magnetization curves point at a small ferromagnetic component present in the compounds with x ≤ 0.25 (Fig. 3). Increase in the dopant content above x = 0.25 leads to a drastic decrease in the magnetic signal accompanied by a notable change in the character of the magnetization curves (Fig. 3). The M(T) and M(H) dependences recorded for the compounds with x > 0.25 indicate the absence of any remanent magnetization and a stabilization of purely collinear antiferromagnetic structure. Careful analysis of the M(H) dependences along with the structural data allowed to clarify the origin of the remanent magnetization and its correlation with the structural parameters of the compounds.

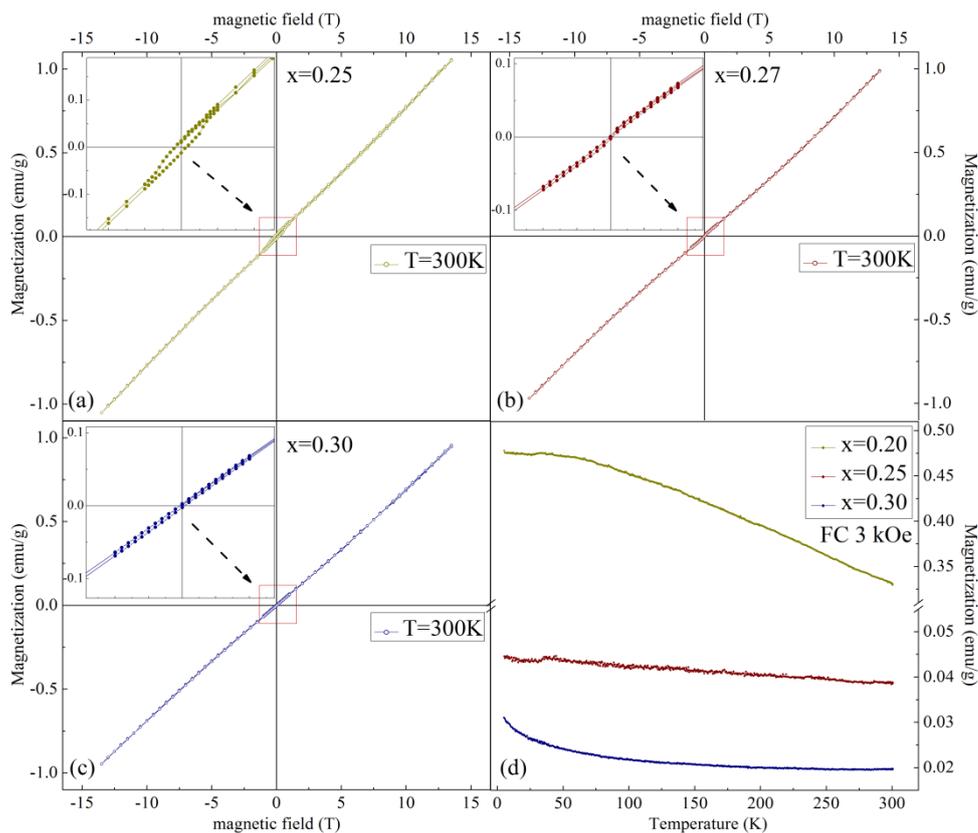

Figure 3. Field (a-c) and temperature (d) dependences of magnetization obtained for the $Bi_{1-x}Ba_xFe_{1-x}Ti_xO_3$ compounds (0.20 ≤ x ≤ 0.30). The insets show magnified parts of the M(H) magnetization curves near the origin.

The value of remanent magnetization of the compounds with x ≤ 0.25 is about 0.02 emu/g at room temperature (Fig. 3) and this value remains nearly constant both in the compounds characterized by single phase rhombohedral structure (x < 0.2) and by the mixture of the rhombohedral and the cubic phases. The compounds with x ≥ 0.27 having dominant cubic structure are characterized by a complete collapse of remanent magnetization which is also characteristic of the compounds with 0.27 ≤ x ≤ 0.4. One should note that a splitting of the reflections (006)$_R$ and (202)$_R$ located at ~39 ° is still distinguishable for the compounds with x=0.27 and x=0.3 (Fig. 1) which confirms the presence of the rhombohedral phase (see Table 1), while the magnetization data show the absence of remanent magnetization. One should also note that in the concentration range 0.25 ≤ x ≤ 0.3, the amount of the rhombohedral phase decreases from 55% down to 10%. Decrease in the amount of the rhombohedral phase leads to a reduction in the value of remanent magnetization, although it cannot explain the drastic decrease in the remanence observed in the compound with x=0.27 having about one third of the rhombohedral phase. Thus, the vanishing of the remanent magnetization in the compounds having a notable amount of the rhombohedral phase (which allows the antisymmetric exchange interaction) as compared to the cubic phase [41] points at the absence of a direct correlation between the type of lattice symmetry (rhombohedral *vs.* cubic) and a non-zero remanence.

It is known that spontaneous magnetization in BiFeO$_3$ based compounds are determined by spin canting originating from the Dzyaloshinskii-Moriya interaction [23, [30]. In the Hamiltonian describing magnetic properties of the BiFeO$_3$ based compounds [42, 43] there are two competitive terms associated with non-collinear magnetism. The first term - $\sum_{ij} K_i(\boldsymbol{\omega}_i - \boldsymbol{\omega}_j) \cdot (\boldsymbol{m}_i \times \boldsymbol{m}_j)$ is related to the oxygen octahedra tilting, the second one - $E_u = -\sum_{ij} C_{ij}(\boldsymbol{u}_i \times \boldsymbol{e}_{ij}) \cdot (\boldsymbol{m}_i \times \boldsymbol{m}_j)$ is associated with electric dipole moments. The non-compensated dipole moments in BiFeO$_3$-based multiferroics with the *R3c* structure are formed via the displacements of ions along the *c*-axis of the hexagonal lattice, thus providing electric polarization along the [001]$_H$ direction. The tilt angle ω as well as the ionic shifts calculated based on the NPD data (Table 1) decrease with an increasing of Ba/Ti content. The most remarkable decrease in the tilt angle ω correlating with the disappearance of remanent magnetization is observed for the compounds with x ≥ 0.27. It should be noted that the magnitude of the tilt angle ω contains both angular distortions associated with a rotation and a tilt of the oxygen octahedra. In the compounds with x ≥ 0.27 the oxygen ions are nearly moved to the coordinate with z$_0$~1 (see Table 1) thus denoting zero value of a tilt of oxygen octahedra while x$_0$ and y$_0$ coordinates of the oxygen ions still indicate oxygen rotation in the basal (*ab*) plane. The obtained results testify a tilt of oxygen octahedra to be the critical structural parameter which determine the value of remanent magnetization.

Another model to explain the evolution of the magnetic properties specific to the compounds within the phase boundary region is based on possible modification of the oxidation state of the iron ions under chemical doping. Based on the neutron diffraction data, one can assume the dominant oxidation state of the iron ions to be 3+, although certain fluctuations are possible due to the local inhomogeneities characteristic for the compounds within the phase coexistence region. This issue was examined by Mössbauer measurements discussed in the next section.

*Mössbauer measurements*

Results of Mössbauer measurements have allowed to estimate an electronic configuration of the iron ions and a symmetry of oxygen octahedra surrounding iron ions. A combination of the data related to the structure and magnetic properties obtained by macroscopic measurements along with the local scale testing performed by Mössbauer spectroscopy have allowed to justify the model describing magnetic properties of the compounds as a function of the dopant content and temperature. Mössbauer spectra recorded for the compounds $Bi_{1-x}Ba_xFe_{1-x}Ti_xO_3$ (0.15 ≤ x ≤ 0.3) are similar to those observed for $BiFeO_3$ only at low temperature (Table 2) [44]. The spectra were refined assuming two sextets (A and B) having slight asymmetry at lower (10 – 200 K) temperatures and the hyperfine field distribution was applied for fitting at room and higher temperatures (Fig. 4, S2, S3). Asymmetry in $BiFeO_3$ Mössbauer spectra was explained by hyperfine field anisotropy related with the alternation of angle between the principal axis of the field gradient tensor and the direction of Fe spin due spiral magnetic structure [45]. Larger amount of the dopant content in the compounds $Bi_{1-x}Ba_xFe_{1-x}Ti_xO_3$ causes a broadening of spectra and decrease in hyperfine field because of chemical dilution of iron subsystem by $Ti^{4+}$ ions. However, no doublet was observed at room temperature as in case of Ref. [46] indicating quite uniform chemical composition. Moreover, a substitution by nonmagnetic Ti ions results in a lowering of exchange energy. However, the cation substitution does not only lower exchange energy but frustrates the spin order. For the $Bi_{1-x}Ba_xFe_{1-x}Ti_xO_3$ compounds, the chemical substitution leads to a distortion of the lattice and results in a change of hyperfine parameters – hyperfine field and quadrupole shift [47] as compared with $BiFeO_3$ (Table 2). Temperature increase causes a transition to paramagnetic state confirming the magnetic transition temperatures estimated based on the neutron diffraction data (Figs. S2, S3).

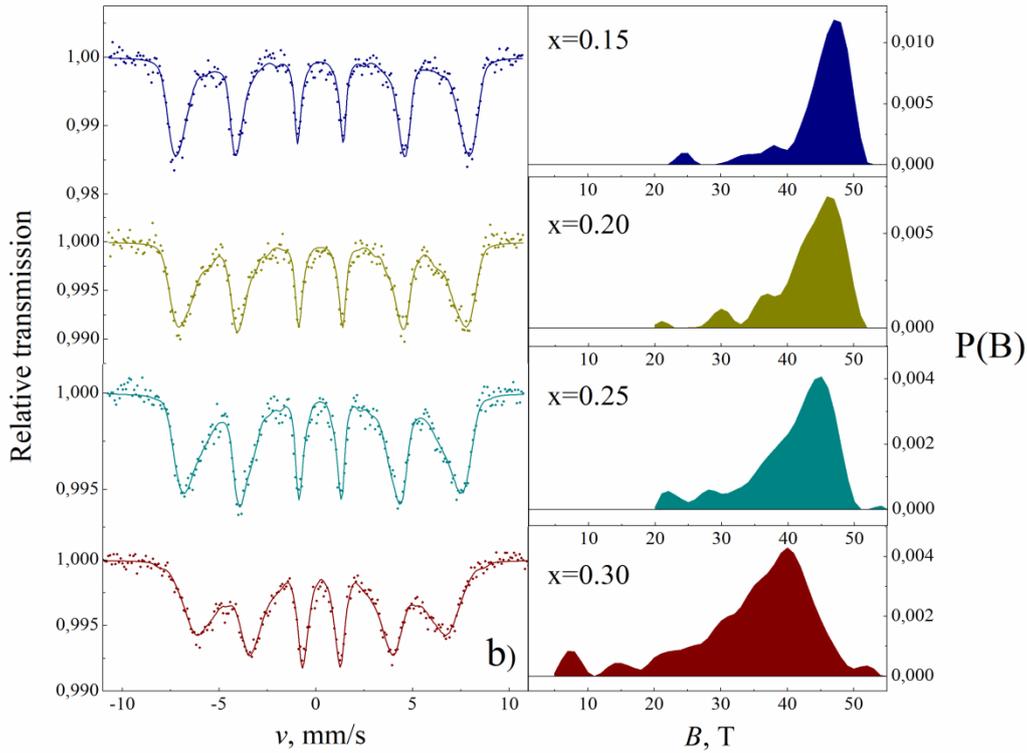

Figure 4. Mössbauer spectra of the compounds $Bi_{1-x}Ba_xFe_{1-x}Ti_xO_3$ with $0.15 \leq x \leq 0.3$ at room temperature (left), hyperfine field distributions P(B) are presented on the right image.

Decrease in quadrupole shift of two sextets refined for the compounds under study can be ascribed to the iron ions having slightly different surrounding associated with the rhombohedral and cubic-like phases (the cubic-like phase is related to the sextets having low value of quadrupole shift). The weight ratio estimated for the sextets only slightly depends on the chemical composition and testifies nearly equal amounts of the averaged respective structural positions. Slight evolution in the ratio of the sextets do not follow the phase ratio estimated by the diffraction methods and can be caused by local character of Mössbauer measurements.

Table 2. Mössbauer parameters relative area, isomer shift $\delta$, quadrupole shift $2\varepsilon$ and hyperfine field $B$ at T~10 K.

| x, % | subspectra | A, % | $\delta$, mm/s | $2\varepsilon$, mm/s | B, T |
|---|---|---|---|---|---|
| 0.15 | A | 53 | 0.50±0.01 | 0.19±0.02 | 54.71±0.12 |
|  | B | 47 | 0.51±0.01 | 0.10±0.03 | 53.01±0.14 |
| 0.20 | A | 50* | 0.51±0.01 | 0.16±0.01 | 54.32±0.05 |
|  | B | 50* | 0.50±0.01 | 0.09±0.02 | 52.64±0.08 |
| 0.25 | A | 49 | 0.51±0.01 | 0.16±0.02 | 54.18±0.11 |
|  | B | 51 | 0.51±0.01 | 0.09±0.02 | 52.40±0.15 |
| 0.30 | A | 50* | 0.49±0.01 | -0.01±0.01 | 53.71±0.07 |
|  | B | 50* | 0.49±0.01 | -0.02±0.01 | 51.77±0.07 |

* fixed

Average hyperfine field <B> and hyperfine fields $B$ of both sextets gradually decrease with the dopant content (Fig. 4, Fig. S3, Table 2) while the values of isomer shift remain nearly constant. The parameters of the Mössbauer spectra confirm 3+ oxidation state of the iron ions regardless the chemical composition of the compounds as well as oxygen octahedra surrounding of the iron ions. Increase in the dopant content above 30 % leads to drastic reduction in the intensity of the Mössbauer spectra which impedes a reliable estimation of the mentioned parameters. The oxidation state and symmetry of oxygen surrounding calculated based on the Mössbauer spectra recorded for the compounds under study are in accordance with the diffraction data and magnetometry results.

4. CONCLUSIONS

The results of diffraction measurements of the compounds $Bi_{1-x}Ba_xFe_{1-x}Ti_xO_3$ ($x \leq 0.40$) indicate that an increase in the dopant concentration leads to a gradual reduction of the rhombohedral distortions. The structure of the compounds with $x = 0.25 - 0.33$ can be refined assuming a coexistence of the rhombohedral and cubic phases; further increase in the dopant content leads to the phase transition to the single phase cubic structure. Analysis of the isothermal dependences of the magnetization as well as neutron diffraction measurements points at the G-type antiferromagnetic structure which is stable in the compounds with $0.15 \leq x \leq 0.4$ in the wide temperature range in spite of the chemical dilution by nonmagnetic Ti ions. The obtained results imply that magnetic properties of the polar phase are strongly dependent on the structural distortions involving the oxygen octahedra tilting which mainly determines the modification of remanent magnetization. The absence of a direct correlation between the type of lattice system and the existence of remanent magnetization observed for the co-doped compounds testifies an inapplicability of the related model used to describe the magnetic properties of $BiFeO_3$ compounds doped with rare-earth ions to those co-substituted by Ba and Ti ions.

ACKNOWLEDGMENTS

This project has received funding from the European Union's Horizon 2020 research and innovation programme under the Marie Skłodowska-Curie grant agreement No 778070 – TransFerr – H2020-MSCA-RISE-2017. The authors acknowledge HZB for the allocation of neutron radiation beamtime and HZB staff for the assistance with neutron diffraction experiments. M.V.S. acknowledges financial support from the Ministry of Science and Higher Education of the Russian Federation within the framework of state support for the creation and development of World-Class Research Centers "Digital biodesign and personalized healthcare" №075-15-2020-926. S.I.L. and D.V.K. acknowledge BRFFR (projects T20R-121 and F20R-123). V.A.K. acknowledges the support provided by national funds from FCT – Fundação para a Ciência e a Tecnologia, I.P., within the projects UIDB/04564/2020 and UIDP/04564/2020.